\documentclass[aps,prb,twocolumn,superscriptaddress]{revtex4-1}
\pdfoutput=1
\newcommand{\figurescale}{1}

\usepackage{bm}
\usepackage{verbatim}
\usepackage{amsmath}
\usepackage[utf8]{inputenc}
\usepackage[pdftex]{graphicx}
\usepackage[separate-uncertainty=true]{siunitx}
\DeclareSIUnit{\rpm}{rpm}
\usepackage[colorlinks=true,urlcolor=blue,linkcolor=blue,citecolor=blue]{hyperref}
\usepackage{xcolor}
\usepackage[T1]{fontenc}
\usepackage{placeins}
\usepackage{nicefrac}
\usepackage{braket}
\usepackage{pdfcomment}
\usepackage{xcolor}
\usepackage[euler]{textgreek}

\usepackage{pbox}
\usepackage{makecell}
\usepackage{hhline}

\definecolor{greentwo}{RGB}{0,180,0}

\begin{document}

\title{Direct exciton emission from atomically thin transition metal dichalcogenide heterostructures near the lifetime limit}
%
\author{J. Wierzbowski}\email{jakob.wierzbowski@wsi.tum.de}
\affiliation{Walter Schottky Institut and Physik Department, Technische Universit\"at M\"unchen, Am Coulombwall 4, 85748 Garching, Germany}
\affiliation{Nanosystems Initiative Munich (NIM), Schellingstr. 4, 80799 Munich, Germany}
\author{J. Klein}
\affiliation{Walter Schottky Institut and Physik Department, Technische Universit\"at M\"unchen, Am Coulombwall 4, 85748 Garching, Germany}
\affiliation{Nanosystems Initiative Munich (NIM), Schellingstr. 4, 80799 Munich, Germany}
\author{F. Sigger}
\affiliation{Walter Schottky Institut and Physik Department, Technische Universit\"at M\"unchen, Am Coulombwall 4, 85748 Garching, Germany}
\author{C. Straubinger}
\affiliation{Walter Schottky Institut and Physik Department, Technische Universit\"at M\"unchen, Am Coulombwall 4, 85748 Garching, Germany}
\author{M. Kremser}
\affiliation{Walter Schottky Institut and Physik Department, Technische Universit\"at M\"unchen, Am Coulombwall 4, 85748 Garching, Germany}
\author{T. Taniguchi}
\affiliation{National Institute for Materials Science, Tsukuba, Ibaraki 305-0044, Japan}
\author{K. Watanabe}
\affiliation{National Institute for Materials Science, Tsukuba, Ibaraki 305-0044, Japan}
\author{U. Wurstbauer}
\affiliation{Walter Schottky Institut and Physik Department, Technische Universit\"at M\"unchen, Am Coulombwall 4, 85748 Garching, Germany}
\affiliation{Nanosystems Initiative Munich (NIM), Schellingstr. 4, 80799 Munich, Germany}
\author{A. W. Holleitner}
\affiliation{Walter Schottky Institut and Physik Department, Technische Universit\"at M\"unchen, Am Coulombwall 4, 85748 Garching, Germany}
\affiliation{Nanosystems Initiative Munich (NIM), Schellingstr. 4, 80799 Munich, Germany}
\author{M. Kaniber}
\affiliation{Walter Schottky Institut and Physik Department, Technische Universit\"at M\"unchen, Am Coulombwall 4, 85748 Garching, Germany}
\affiliation{Nanosystems Initiative Munich (NIM), Schellingstr. 4, 80799 Munich, Germany}
\author{K. M\"uller}
\affiliation{Walter Schottky Institut and Physik Department, Technische Universit\"at M\"unchen, Am Coulombwall 4, 85748 Garching, Germany}

\author{J. J. Finley}
\affiliation{Walter Schottky Institut and Physik Department, Technische Universit\"at M\"unchen, Am Coulombwall 4, 85748 Garching, Germany}
\affiliation{Nanosystems Initiative Munich (NIM), Schellingstr. 4, 80799 Munich, Germany}
%
%
\date{\today}
%
%

\begin{abstract}
\textbf{We demonstrate the reduction of the inhomogeneous linewidth of the free excitons in atomically thin transition metal dichalcogenides (TMDCs) MoSe$_{2}$, WSe$_{2}$ and MoS$_{2}$ by encapsulation within few nanometer thick hBN.
Encapsulation is shown to result in a significant reduction of the 10K excitonic linewidths down to $\sim\SI{3.5}{\milli\electronvolt}$ for n-MoSe$_{2}$, $\sim\SI{5.0}{\milli\electronvolt}$ for p-WSe$_{2}$ and $\sim\SI{4.8}{\milli\electronvolt}$ for n-MoS$_{2}$.
Evidence is obtained that the hBN environment effectively lowers the Fermi level since the relative spectral weight shifts towards the neutral exciton emission in n-doped TMDCs and towards charged exciton emission in p-doped TMDCs.
Moreover, we find that fully encapsulated MoS$_{2}$ shows resolvable exciton and trion emission even after high power density excitation in contrast to non-encapsulated materials.
Our findings suggest that encapsulation of mechanically exfoliated few-monolayer TMDCs within nanometer thick hBN dramatically enhances optical quality, producing ultra-narrow linewidths that approach the homogeneous limit.}
\end{abstract}
%
%
\maketitle
%
%
\section{Introduction}

In the group of atomically thin two-dimensional (2D) materials the transition metal dichalcogenides MoS$_{2}$, MoSe$_{2}$, WS$_{2}$ and WSe$_{2}$ reveal fascinating photophysical properties owing to their direct gap and strong light-matter interactions\cite{Splendiani.2010,Mak.2010}. The weak dielectric screening results in emission dominated by excitonic processes, with exciton binding energies on the order of several hundred $\SI{}{\milli\electronvolt}$ \cite{He.2014,Ugeda.2014} that follow a non-hydrogenic Rydberg series\cite{Chernikov.2014}.
However, in the vast majority of reports to date the linewidths of the free excitons exhibit significant inhomogeneous broadening. This is typically attributed to the local spatial inhomogeneity of the substrate, adsorbed atoms and molecules on the surface due to the large surface-to-volume ratio and different doping and dielectric screening conditions that are highly sensitive to the choice of substrate.
Broad linewidths of the exciton emission of $\sim\SI{50}{\milli\electronvolt}$ for MoS$_{2}$,\cite{Splendiani.2010, Mak.2010, Sercombe.2013} $\sim\SI{75}{\milli\electronvolt}$ for WS$_{2}$,\cite{Zhao.2013} $\sim\SI{5}{\milli\electronvolt}$ for MoSe$_{2}$ \cite{Ross.2013} and $\sim\SI{10}{\milli\electronvolt}$ for WSe$_{2}$ \cite{Zhao.2013, Jones.2013} have been reported in photoluminescence experiments, while time-domain spectroscopy \cite{Moody.2015, Jakubczyk.2016, Dey.2016} and recent theory \cite{Selig.2016} report homogeneously broadened luminescence linewidths in the range of $\sim$ $2-\SI{6}{\milli\electronvolt}$ depending on the material system.
The healing of sulfur defects using sulfuric superacids increases the optical quantum yield and reduces the linewidths at room temperature\cite{amani2015near,amani2016recombination,cadiz2016well} from $\sim\SI{70}{\milli\electronvolt}$ to $\sim\SI{55}{\milli\electronvolt}$. However, low temperature studies of treated MoS$_{2}$ monolayers\cite{cadiz2016well} show that the linewidths still remain in the order of $\sim\SI{15}{\milli\electronvolt}$.
Very recently, it has been shown that MoS$_{2}$ is particularly sensitive to photoinduced irreversible changes resulting in broad luminescence from overlapping neutral and charged exciton emission\cite{Cadiz.2016}. Measurements performed using ultra-low excitation power densities reveal distinct peaks for neutral and charged excitons with linewidths of $\sim\SI{15}{\milli\electronvolt}$ for MoS$_{2}$ similar to Se-based TMDCs\cite{Cadiz.2016}.

In this letter, we present an optical study of TMDCs encapsulated within hBN and demonstrate that encapsulation leads to a significant reduction of the linewidth observed in photoluminescence experiments, towards the radiative limit.
We systematically probe modifications in the luminescence linewidth after each stacking step and extract key parameters such as the exciton peak position, relative intensities of exciton and trion recombination and peak linewidths.
We also show that annealing of the heterostructure improves the spatial homogeneity of the TMDC and, thus, of the observed luminescence.
From our results, we make three major observations upon hBN encapsulation: (i) the linewidths of free excitons are significantly reduced down to a few $\SI{}{\milli\electronvolt}$ approaching the homogeneous linewidth limit, (ii) the surface is protected, preventing samples against irreversible photoinduced spectral changes and (iii) encapsulation lowers the Fermi level, reducing emission from negatively charged excitons in MoSe$_{2}$, while increasing the emission from positively charged excitons in WSe$_{2}$.

%
\begin{figure*}[!ht]
\scalebox{\figurescale}{\includegraphics[width=1\linewidth]{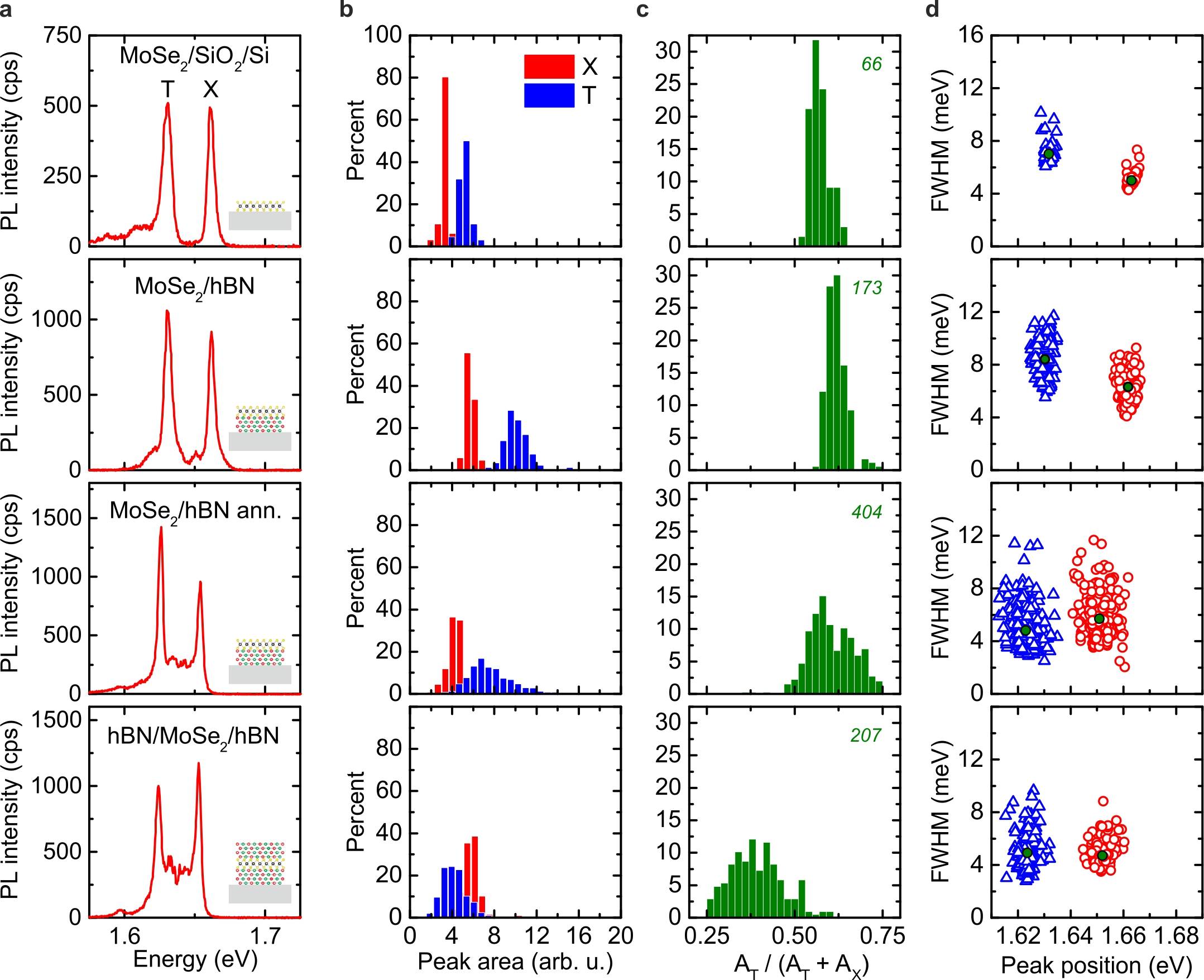}}
\renewcommand{\figurename}{Fig.}
\caption{\label{fig1}
	\textbf{MoSe$_2$ photoluminescence spectra and statistics.}
(a) Typical low-temperature ($\SI{10}{\kelvin}$) \textmu-PL spectra of MoSe$_{2}$ on SiO$_{2}$, MoSe$_{2}$ on hBN, MoSe$_{2}$ on hBN after annealing and MoSe$_{2}$ encapsulated between hBN. Emission is observed from the neutral (X) and charged exciton (T) transitions.
(b) Histogram of the peak areas of X (red, $A_{X}$) and T (blue, $A_{T}$).
(c) Corresponding relative spectral weight $R = A_{T}/(A_{T} + A_{X})$. The green italic number represents the number of fitted spectra used for the histograms.
(d) Correlated distribution of Lorentzian linewidths and corresponding peak positions of X (red circles) and charged (blue triangles) exciton. The green circles and triangles denote the corresponding mean values.
}
\end{figure*}
%

\section{Results and discussion}

\subsection{Sample fabrication and \textmu-PL measurements}

The monolayer TMDCs studied in this letter are mechanically exfoliated onto degenerately n-doped Si substrates covered with a \SI{285}{\nano\metre} thick layer of wet-thermally grown SiO$_{2}$.
The heterostructures are stacked using the dry viscoelastic transfer method\cite{CastellanosGomez.2014}, whereby we iteratively stacked hBN/TMDC/hBN onto the Si/SiO$_{2}$ substrate. The hBN layer thicknesses range from $\SI{10}{\nano\meter}$ to $\SI{70}{\nano\meter}$ (AFM measurements). After stacking, the heterostructures were annealed at $\SI{150}{\celsius}$ for $\SI{20}{\minute}$ to remove water and polymer accumulated into bubbles and improve the sample homogeneity (see supporting information).\newline
All photoluminescence (PL) experiments were performed using a confocal microscope at $\SI{10}{\kelvin}$. The continuous-wave excitation energy was kept at $\SI{2.33}{\electronvolt}$ (Nd:YAG) and an ex\-ci\-ta\-tion po\-wer den\-si\-ty of $\SI{0.66}{\kilo\W\per\centi\meter\squared}$, unless otherwise noted. The spatial mode field diameter of the focal spot ($\nicefrac{1}{e^{2}}$ contour) was $\sim\SI{1.1}{\micro\meter}$. The detected light was filtered with a steep fluorescence filter with a trans\-mis\-sion cut-on energy $\SI{11.7}{\milli\electronvolt}$ below the laser excitation energy.

\subsection{Photoluminescence of encapsulated MoSe$_{2}$}
\label{subsec:mose2}

To probe the impact of the proximal substrate and explore the benefits of hBN encapsulation, we performed spatially resolved PL measurements and statistically analyze the emission spectra at different positions on the sample surface.
Note, in our analysis we disregard spectra recorded from the edge of the flake and obviously damaged parts of the sample, as identified by conventional light microscopy. From the measurements, we extract the peak positions, full widths at half maximum linewidths (FWHM) and relative intensities of the neutral exciton (X) and charged trion (T) by fitting with Lorentzian peaks.
Fig.~\ref{fig1} compares examples of spectra of MoSe$_{2}$ on SiO$_{2}$, MoSe$_{2}$ on hBN, MoSe$_{2}$ on hBN after annealing and MoSe$_{2}$ sandwiched between hBN and after annealing. The corresponding statistical analysis of peak position, exciton linewidth and peak area for the different MoSe$_{2}$/substrate configurations are shown in Fig.~\ref{fig1}b-d, respectively. Note that in order to obtain the best comparison, in the case of MoSe$_{2}$ on hBN we scan the same area after subsequent steps of stacking and annealing to trace the impact of the encapsulation on the spectral evolution.
A typical spectrum recorded from MoSe$_{2}$ on SiO$_{2}$ is presented in Fig.~\ref{fig1}a (top). It exhibits pronounced emission from trions, typically attributed to extrinsic effects such as doping from the substrate, mediated through trap states \cite{Mak.2012, Sercombe.2013, Ross.2013, Jones.2013} and intrinsic doping resulting from chalcogen vacancies and adsorbates that are reported to occur in mechanically exfoliated flakes\cite{Komsa.2012, Hong.2015}. We obtain a qualitative measure of the doping by analyzing the areas of the neutral and charged exciton $A_{X}$ and $A_{T}$ and their relative spectral weight $R = A_{T}/(A_{T} + A_{X})$.
Fig.~\ref{fig1}b shows the peak areas, while the corresponding relative spectral weights are presented in Fig.~\ref{fig1}c. The emission intensity for MoSe$_{2}$ on SiO$_{2}$ is higher for trions than for neutral excitons which is reflected by values of W > 0.5 in Fig.~\ref{fig1}c.
This remains unchanged when MoSe$_{2}$ is stacked on top of $\sim\SI{14}{\nano\meter}$ thick hBN, and also after annealing that only results in a slightly decreased total peak area. However, fully encapsulated MoSe$_{2}$ exhibits a higher X peak-area compared to T with W < 0.5. This behaviour is indicative of a lowering of the Fermi level in the crystal inhibiting trion formation. 
This effect is strongest in the fully encapsulated configuration. Since the MoSe$_{2}$ is exposed to ambient conditions during and after fabrication in the previous configurations, the TMDC surfaces are free to physisorption of ambient molecules \cite{tongay2013broad,miller2015photogating}, most likely H$_{2}$O due to its polarity. Thus, we assume that the impact of the hBN substrate is reduced due to frozen adsorbates possibly at defects such as selenium vacancies on the TMDC surface.\newline
Changes in the dielectric environment and doping can further influence the exciton peak positions and the linewidths (Fig.~\ref{fig1}d)\cite{Komsa.2015}. Here, we directly correlate peak positions and linewidths. The exfoliated MoSe$_{2}$ on SiO$_{2}$ shows exciton peak positions of $P_{X} = \SI{1663.1\pm1.2}{\milli\electronvolt}$ and $P_{T} = \SI{1631.8\pm1.3}{\milli\electronvolt}$ with a binding energy of $E_{T} \sim\SI{31}{\milli\electronvolt}$ which is typically observed in literature\cite{Ross.2013,wang2015polarization}. Stacking MoSe$_{2}$ on hBN results only in a slight redshift by $\Delta E \sim\SI{-2}{\milli\electronvolt}$ and a slightly broader distribution as can be seen in Fig.~\ref{fig1}b. This redshift is consistent with recent calculations\cite{Komsa.2015} and measurements\cite{lin2014dielectric} considering the change in the refractive index of the substrate from n$_{\text{SiO}_{2}} = 1.457$ to n$_{\text{hBN}}=2.2$ at the neutral exciton resonance\cite{malitson1965interspecimen,gorbachev2011hunting}.\newline
Annealing results in an additional redshift, and a total shift to a lower energy by $\Delta E \sim\SI{-12}{\milli\electronvolt}$ compared to pristine MoSe$_{2}$. This is accompanied by a much broader distribution of peak positions.
The sandwiched and annealed MoSe$_{2}$ structure exhibits the strongest redshift of $\Delta E \sim\SI{-21}{\milli\electronvolt}$.
Yet, the statistical spread of the peak position distribution is significantly reduced from $s_{X}=\SI{6.8\pm0.1}{\milli\electronvolt}$ to $s_{X}=\SI{2.8\pm0.1}{\milli\electronvolt}$ (see supporting information), as depicted in the bottom panel in Fig. \ref{fig1}d.
Moreover, the trion binding energy decreases from $E_{X}-E_{T}=\SI{31\pm3}{\milli\electronvolt}$ to $\SI{28\pm3}{\milli\electronvolt}$ after the annealing step possibly resulting from the modification of the dielectric environment and a change in extrinsic doping\cite{Komsa.2015}.
In general, we observe that the symmetric dielectric hBN environment of the MoSe$_{2}$ flake combined with annealing results in the sharpest distribution of emission energies, indicative of the highest homogeneity within the MoSe$_{2}$ flake.
The statistical analysis of the linewidths for MoSe$_{2}$ on SiO$_{2}$ reveals average values of $w_{X} = \SI{5.0\pm0.5}{\milli\electronvolt}$ and $w_{T} = \SI{7.0\pm0.8}{\milli\electronvolt}$ for X and T excitons, respectively.
Stacking MoSe$_{2}$ on hBN results in significantly higher linewidths of $w_{X}=\SI{6.3\pm1.0}{\milli\electronvolt}$ and $w_{T} = \SI{8.4\pm1.3}{\milli\electronvolt}$ with a much broader variation in obtained values. Annealing reduces the linewidth to $w_{X} = \SI{5.7\pm1.5}{\milli\electronvolt}$ and $w_{T} = \SI{4.8\pm1.5}{\milli\electronvolt}$ while capping with hBN further reduces the X linewidth to $w_{X} = \SI{4.7\pm0.9}{\milli\electronvolt}$, keeping the T linewidth at $w_{T} = \SI{4.9\pm1.3}{\milli\electronvolt}$. Interestingly, annealing reveals much higher variance of values which is significantly narrowed upon capping. However, for investigating the linewidths not only the average values are important but also the lowest values obtained. Importantly, for MoSe$_{2}$ encapsulated in hBN we observe values as low as $w_{X} \sim\SI{3.5}{\milli\electronvolt}$,
almost reaching the homogeneous linewidths recently reported in time-resolved four-wave-mixing experiments\cite{Jakubczyk.2016, Dey.2016} and theoretical calculations\cite{Selig.2016} of $w_{X}\sim\SI{2.1}{\milli\electronvolt}$, $w_{X}\sim\SI{3.4}{\milli\electronvolt}$ and $w_{X}\sim\SI{5.5}{\milli\electronvolt}$, for lattice temperatures of $T=\SI{6}{\kelvin}\text{,~} \SI{5}{\kelvin} \text{~and~} \SI{10}{\kelvin}$, respectively\newline
With the dependence $\gamma_{rad}\propto1/n_{\text{Substrate}}$ for the radiative linewidth broadening\cite{knorr1996theory,Selig.2016}, changing the substrate material reduces $\gamma_{rad}$ by a factor of $n_{\text{SiO$_{2}$}}/ n_{\text{hBN}} \approx 0.66$. This then produces a radiative rate which would be quantitatively consistent with the narrowest linewidths measured in our study. Beside radiative broadening, primarily exciton-phonon coupling has been identified as broadening mechanism\cite{Selig.2016}. Moreover, we attribute the observed remaining broadening of the linewidth to spatial inhomogeneities of the TMDC as a result of the exfoliation procedure and residual polymer bubbles between the interfaces of the monolayer crystal and the surrounding hBN.

\subsection{Photoluminescence of encapsulated WSe$_{2}$}
\begin{figure*}[t!]
	\scalebox{\figurescale}{\includegraphics[width=1\linewidth]{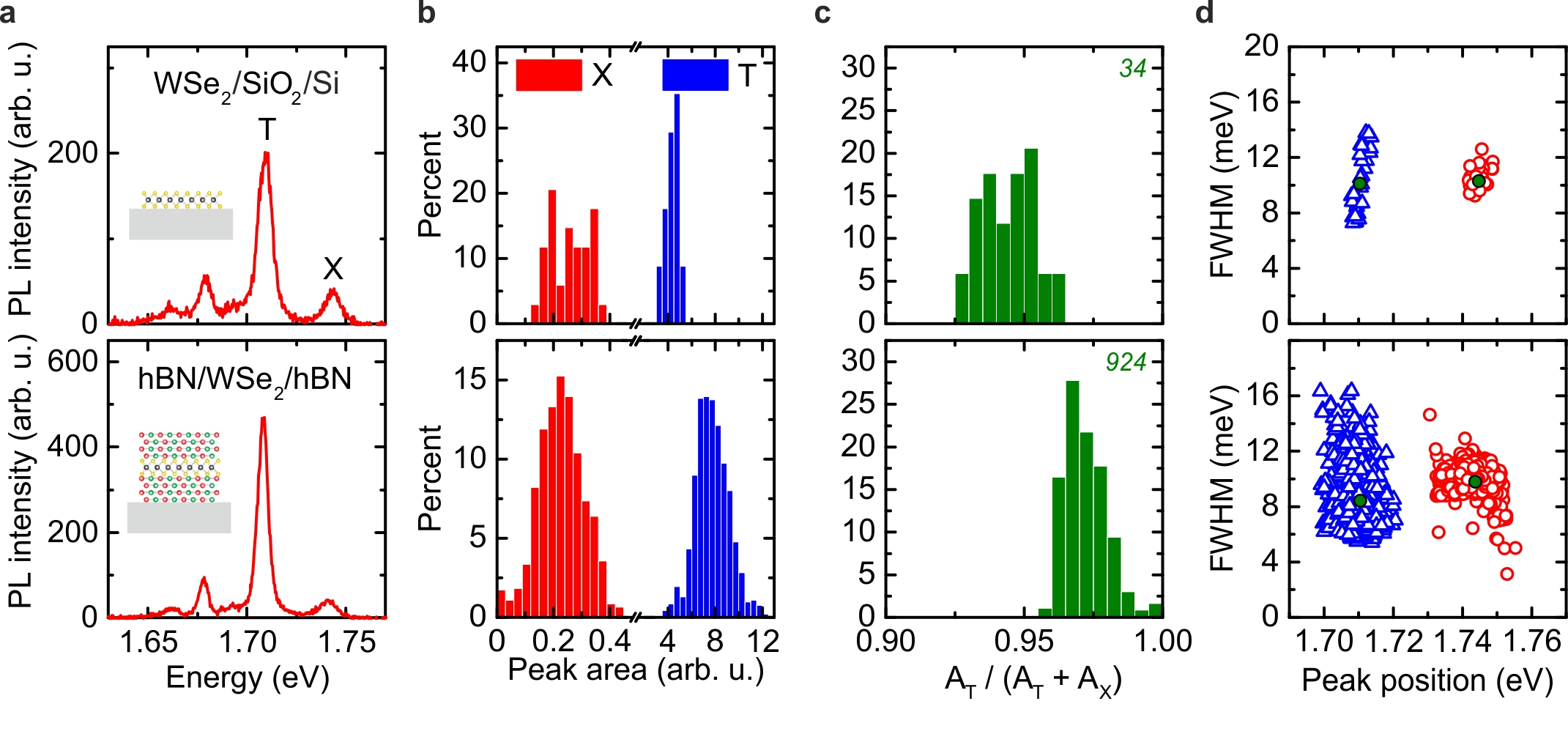}}
	\renewcommand{\figurename}{Fig.}
	\caption{\label{fig2}
		\textbf{WSe$_2$ photoluminescence spectra and statistics.}
		(a) Typical low-temperature ($\SI{10}{\kelvin}$) \textmu-PL spectrum WSe$_{2}$ on SiO$_{2}$ and encapsulated within hBN featuring emission from the neutral (X) and charged exciton (T).
		(b) Histogram of peak areas of X (red, $A_{X}$) and T (blue, $A_{T}$).
		(c) Corresponding relative spectral weight $R = A_{T}/(A_{T} + A_{X})$. The green italic number represents the fitted spectra used for the histograms.
		(d) Correlated distribution of Lorentzian linewidths and corresponding peak positions of X (red circles) and charged (blue triangles) exciton. The green circles and triangles denote the corresponding mean values.
	}
\end{figure*}
We repeated the fabrication scheme and optical experiments discussed above for MoSe$_{2}$ with WSe$_{2}$. Since we found the most significant improvement in optical quality for TMDCs that are fully encapsulated in hBN, we compare only the two cases of WSe$_{2}$ on SiO$_{2}$ and WSe$_{2}$ encapsulated in hBN after annealing. Typical spectra for WSe$_{2}$ on hBN are presented in Fig.~\ref{fig2}a. The relative spectral weights $R$ are shown in Fig.~\ref{fig2}b.
Comparing the relative peak areas of the neutral and charged excitons (relative spectral weight in section \ref{subsec:mose2}), results in a trend opposite to that for MoSe$_{2}$.
For WSe$_{2}$, the relative intensity of the charged exciton increases by a factor of two upon encapsulation with hBN. We explain this trend by the difference in intrinsic doping of TMDCs present in our experiments. The MoSe$_{2}$ crystal employed in this work is n-doped, which results in a negatively charged exciton. In contrast, the WSe$_{2}$ is p-doped, resulting in emission from positively charged excitons (see supporting information). Thus, the hBN encapsulation effectively shifts the Fermi level in the TMDC to lower values, enabling a higher positively charged exciton formation rate.\newline
Upon encapsulation, we observe a reduction of the neutral exciton emission linewidth from $\SI{10.3\pm0.7}{\milli\electronvolt}$ to $\SI{9.8\pm1.4}{\milli\electronvolt}$, whilst the trion emission linewidth reduces from $\SI{10.1\pm2.1}{\milli\electronvolt}$ to $\SI{8.4\pm1.9}{\milli\electronvolt}$. However, this effect is accompanied by a higher overall spread in the linewidth distribution for the capped material. A similar trend is observed for the distribution of peak positions. However, here only a slight redshift is observed.
Notably, we observe linewidths as low as $w_{X} \sim \SI{5}{\milli\electronvolt}$ for the neutral exciton and $w_{T} \sim \SI{5.5}{\milli\electronvolt}$.
Recent four-wave-mixing measurements\cite{Moody.2015,Dey.2016} and theoretical work\cite{Selig.2016} report and predict homogeneous linewidths of $w_{X}\sim\SI{6.1}{\milli\electronvolt} $, $w_{X}\sim\SI{4.72}{\milli\electronvolt}$ and $w_{X}\sim\SI{6.5}{meV}$, respectively.\newline
Optimised stacking processes, reducing bubble formation and wrinkling of the 2D materials could lead to desired purely lifetime broadened emission of the TMDCs.

\subsection{Photoluminescence of encapsulated MoS$_{2}$}

\begin{figure}[!ht]
	\scalebox{\figurescale}{\includegraphics[width=1\linewidth]{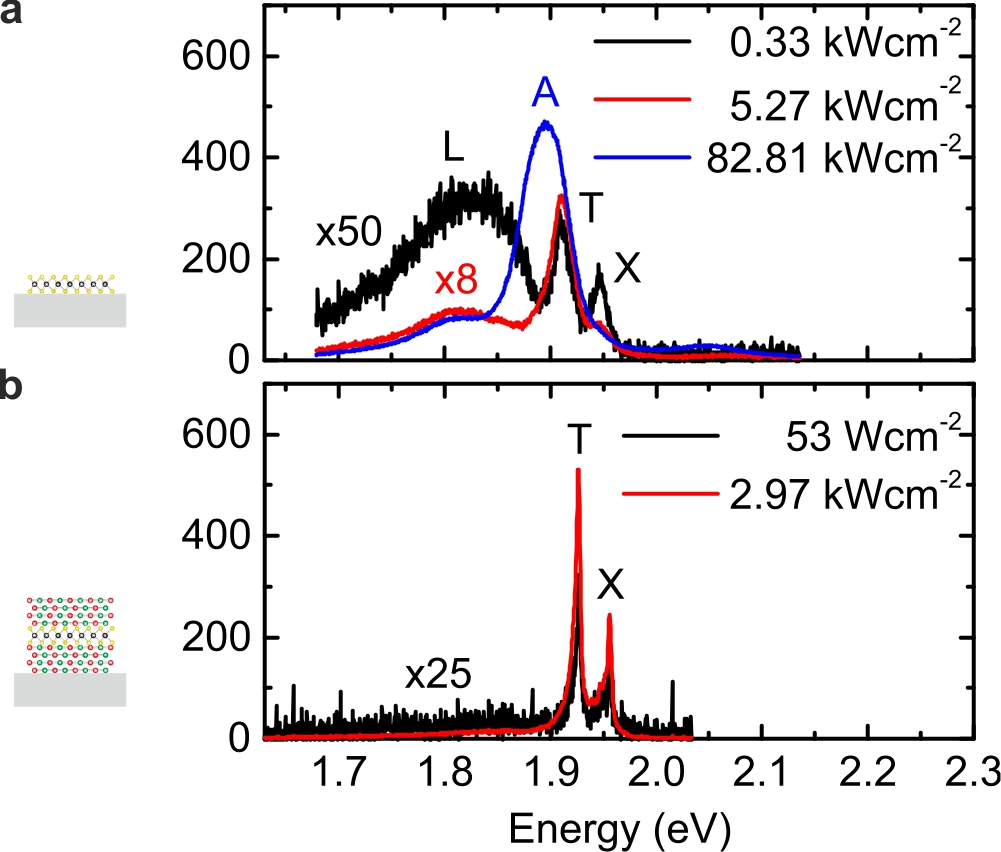}}
	\renewcommand{\figurename}{Fig.}
	\caption{\label{fig3}
		\textbf{Power dependent MoS$_2$ photoluminescence spectra.}
		(a) Typical \textmu-PL spectrum of MoS$_{2}$ on SiO$_{2}$ for a low (black) moderate (red) and high (blue) excitation power featuring the A-peak (blue spectrum) neutral and charged exciton emission and the L-peak at lower energies.
		(b) Typical \textmu-PL spectrum of hBN encapsulated MoS$_{2}$ for a low (black) and high (red) excitation power reveals sharp neutral and charged exciton emission and no emission from the L-peak.
	}
\end{figure}

In addition to the Se based TMDCs, we also applied our encapsulation scheme to MoS$_{2}$ which in past experiments showed comparatively broad emission from the A-exciton \cite{Splendiani.2010, Mak.2010, Sercombe.2013}. This is attributed to inhomogeneous broadening of the emission from neutral and charged excitons that is typically so large that the two peaks are not resolved.
Typical PL from MoS$_{2}$ exfoliated on SiO$_{2}$ is presented in Fig.~\ref{fig3}a. For very low excitation power densities of $\SI{0.33}{\kilo\watt\per\centi\meter\squared}$, the spectrum (black curve) reveals emission from the neutral exciton X at $\SI{1947.4\pm0.3}{\milli\electronvolt}$, charged excitons T at $\SI{1910.7\pm0.3}{\milli\electronvolt}$ and pronounced emission from the low energy L-peak is observed located $\sim\SI{100}{\milli\electronvolt}$ below X. This broad emission is attributed to defect-related exciton emission\cite{Splendiani.2010, Mak.2010, Korn.2011}. Upon increasing the excitation power density to $\SI{5.27}{\kilo\watt\per\centi\meter\squared}$ (red curve in Fig.~\ref{fig3}a) the emission from the neutral exciton vanishes while charged exciton emission dominates. Meanwhile the emission from the L-peak saturates, and its contribution reduces compared to the charged exciton emission. When further increasing the excitation power density to values as high as $\SI{83}{\kilo\watt\per\centi\meter\squared}$ (blue curve in Fig.~\ref{fig3}a), the emission merges to the broad A-exciton peak normally observed in luminescence studies of MoS$_{2}$ with a linewidth of $w_{A} \sim\SI{53.6\pm0.8}{\milli\electronvolt}$\cite{Mak.2010}. Note that these photoinduced changes in the form of the PL spectrum in our studies were found to be irreversible, consistent with recent findings\cite{Cadiz.2016}.
For the lowest excitation power densities investigated, the neutral and charged excitons exhibit linewidths of $w_{X} \sim\SI{14.7\pm0.7}{\milli\electronvolt}$ and $w_{T} \sim\SI{23.4\pm0.8}{\milli\electronvolt}$. Here, a full statistical analysis was not possible due to the photoinduced changes in the optical spectra.
In strong contrast, encapsulation of MoS$_{2}$ and annealing significantly enhances the optical emission properties. The PL (Fig.~\ref{fig3}b) exhibits bright emission from free excitons. The neutral exciton at $\SI{1955.8\pm0.5}{\milli\electronvolt}$ and the trion emission at $\SI{1926.2\pm0.5}{\milli\electronvolt}$ is now blue shifted by $\SI{8.4\pm1}{\milli\electronvolt}$ and $\SI{15.5\pm1}{\milli\electronvolt}$ compared to the MoS$_{2}$ on SiO$_{2}$ configuration, respectively.
By comparing the bare monolayer on SiO$_{2}$ at $\sim\SI{5}{\kilo\watt\per\centi\meter\squared}$ to the encapsulated MoS$_{2}$ at $\sim\SI{3}{\kilo\watt\per\centi\meter\squared}$ (red curves in Fig.\ref{fig3}a and b), we observe that the relative spectral weight strongly shifts from $\sim0.94$ towards lower values of $\sim0.75$. This behaviour of the relative spectral weight of the charged trion emission indicates that the hBN effectively lowers the Fermi level in the MoS$_{2}$.
This blueshift is accompanied by a strong decrease in X and T linewidths down to  $w_{X} \sim\SI{4.8\pm1}{\milli\electronvolt}$ and $w_{T} \sim\SI{6.8\pm0.9}{\milli\electronvolt}$, consistent with recent work by Dey \emph{et al.}\cite{Dey.2016} reporting a homogeneous linewidth of $w_{X}\sim\SI{6.6}{\milli\electronvolt}$ in time-resolved four-wave-mixing measurements.\newline
Importantly, we observe no emission from the L-peak indicative of defects and adsorbates\cite{Splendiani.2010, Mak.2010, Korn.2011} for the fully encapsulated sample.  Such features are observed for all other sample configurations, further highlighting the importance of surface protection.
Furthermore, both exciton species are well resolved and we observe no photoinduced changes even for the highest excitation power ($\SI{83}{\kilo\watt\per\centi\meter\squared}$) used in our experiments.

%

\section{Conclusion}

In summary, we have investigated the impact of hBN encapsulation on the optical properties of several TMDs through statistically analyzing low temperature photoluminescence experiments.
Encapsulation distinctly reduces exciton linewidths and further shifts the Fermi level of the TMDCs. Moreover, surface protection especially enhances the optical quality of MoS$_{2}$, resulting in very clean spectra and revealing sharp emission from neutral and charged exciton without the presence of any irreversible photoinduced changes. Our findings suggest that encapsulation of TMDCs is essential for accessing the interesting photophysical properties of MoS$_{2}$ and enables more sophisticated future optoelectronic devices.

During the writing of this manuscript we recognised related work reported by Cadiz \emph{et al}. \cite{Cadiz.2017} and Ajayi \emph{et al}. \cite{Ajayi.2017}.

%
%
\section{Acknowledgements}
We gratefully acknowledge financial support from ExQM PhD programme of the Elite Network of Bavaria, the German Excellence Initiative via the Nanosystems Initiative Munich (NIM), the Deutsche Forschungsgemeinschaft (DFG) through the TUM International Graduate School of Science and Engineering (IGSSE) and the International Max Planck Research School for Quantum Science and Technology (IMPRS-QST).\newline

\section{Author contributions}
\noindent J.W. and J.K. contributed equally.\newline
 
\section{Abbreviations}
TMDC, transition metal dichalcogenides; \textmu-PL, Micro-photoluminescence.

%
\section{Additional information}
\subsection{Supplementary Information}
Supporting information accompanies this paper.

\subsection{Competing financial interests}
The authors declare no competing financial interests.

%
%

\bibliographystyle{apsrev4-1}
\bibliography{MainBibliography}

\end{document}


\title{Supporting information: Direct exciton emission from atomically thin transition metal dichalcogenide heterostructures near the lifetime limit}
\date{\today}
\author{J. Wierzbowski}\email{jakob.wierzbowski@wsi.tum.de}
\affiliation{Walter Schottky Institut and Physik Department, Technische Universit\"at M\"unchen, Am Coulombwall 4, 85748 Garching, Germany}
\affiliation{Nanosystems Initiative Munich (NIM), Schellingstr. 4, 80799 Munich, Germany}
\author{J. Klein}
\affiliation{Walter Schottky Institut and Physik Department, Technische Universit\"at M\"unchen, Am Coulombwall 4, 85748 Garching, Germany}
\affiliation{Nanosystems Initiative Munich (NIM), Schellingstr. 4, 80799 Munich, Germany}
\author{F. Sigger}
\affiliation{Walter Schottky Institut and Physik Department, Technische Universit\"at M\"unchen, Am Coulombwall 4, 85748 Garching, Germany}
\author{C. Straubinger}
\affiliation{Walter Schottky Institut and Physik Department, Technische Universit\"at M\"unchen, Am Coulombwall 4, 85748 Garching, Germany}
\author{M. Kremser}
\affiliation{Walter Schottky Institut and Physik Department, Technische Universit\"at M\"unchen, Am Coulombwall 4, 85748 Garching, Germany}
\author{T. Taniguchi}
\affiliation{National Institute for Materials Science, Tsukuba, Ibaraki 305-0044, Japan}
\author{K. Watanabe}
\affiliation{National Institute for Materials Science, Tsukuba, Ibaraki 305-0044, Japan}
\author{U. Wurstbauer}
\affiliation{Walter Schottky Institut and Physik Department, Technische Universit\"at M\"unchen, Am Coulombwall 4, 85748 Garching, Germany}
\affiliation{Nanosystems Initiative Munich (NIM), Schellingstr. 4, 80799 Munich, Germany}
\author{A. W. Holleitner}
\affiliation{Walter Schottky Institut and Physik Department, Technische Universit\"at M\"unchen, Am Coulombwall 4, 85748 Garching, Germany}
\affiliation{Nanosystems Initiative Munich (NIM), Schellingstr. 4, 80799 Munich, Germany}
\author{M. Kaniber}
\affiliation{Walter Schottky Institut and Physik Department, Technische Universit\"at M\"unchen, Am Coulombwall 4, 85748 Garching, Germany}
\affiliation{Nanosystems Initiative Munich (NIM), Schellingstr. 4, 80799 Munich, Germany}
\author{K. M\"uller}
\affiliation{Walter Schottky Institut and Physik Department, Technische Universit\"at M\"unchen, Am Coulombwall 4, 85748 Garching, Germany}
\author{J. J. Finley}
\affiliation{Walter Schottky Institut and Physik Department, Technische Universit\"at M\"unchen, Am Coulombwall 4, 85748 Garching, Germany}
\affiliation{Nanosystems Initiative Munich (NIM), Schellingstr. 4, 80799 Munich, Germany}

\maketitle

\section{TMDC crystals}
\noindent The employed n-MoSe$_{2}$ and p-WSe$_{2}$ crystals were purchased at HQGraphene (www.hqgraphene.com), Netherlands and the MoS$_{2}$ was purchased at SPI Supplies (www.2spi.com), USA. The doping information is provided on the corresponding websites.

\subsection{Effects of annealing}
\begin{wrapfigure}{r}{0.45\textwidth}
	\begin{center}
		\includegraphics[width=0.45\textwidth]{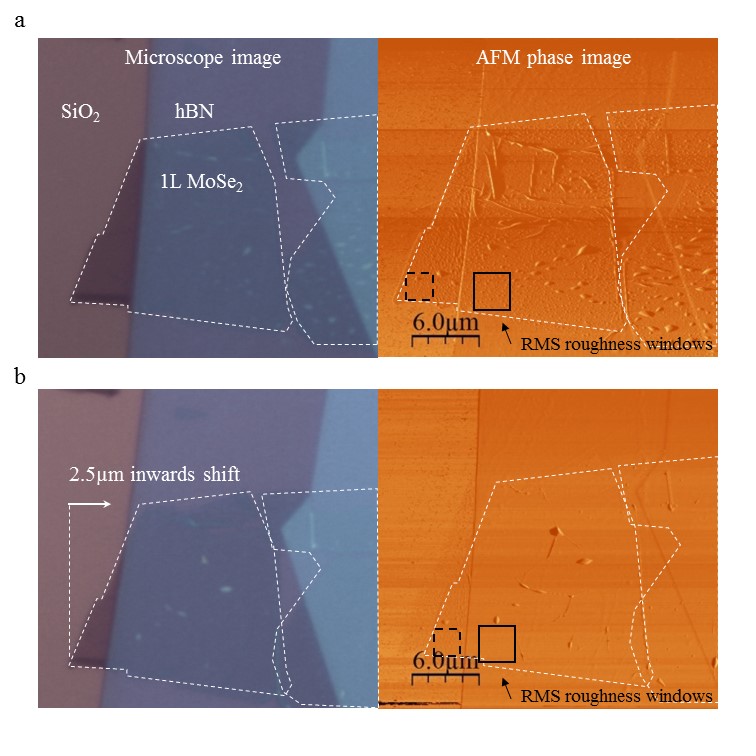}
	\end{center}
	\renewcommand{\figurename}{Fig.}
	\caption{
		\textbf{Optical microscope and AFM phase images of MoSe$_{2}$.}
		Annealing effects of MoSe$_{2}$ on hBN. (a) Microscope (left panel) and AFM phase images (right panel) of MoSe$_{2}$ before annealing. The white dashed regions show the MoSe$_{2}$ monolayer crystal partially covering SiO$_{2}$ and hBN regions. The solid (dashed) black $3\times\SI{3}{\micro\meter\squared}$ ($2\times\SI{2}{\micro\meter\squared}$) squares indicates a region for RMS roughness analysis. (b) MoSe$_{2}$ monolayer after annealing at $\SI{150}{\degreeCelsius}$. The solid white arrow indicates a lateral shift of the monolayer material into the hBN region. The phase images are shown to emphasise the morphology. The RMS roughness analysis was performed on the actual height profiles.
	}
	\label{figAFM}
\end{wrapfigure}

\noindent In order to illustrate the influences of annealing, we present in figure \ref{figAFM} two microscope and AFM phase images\cite{horcas2007wsxm} of a region of a MoSe$_{2}$/hBN structure, before and after the annealing process. We employ the phase images due to contrast reasons. The RMS analysis was carried out on the actual height profiles.

Figure \ref{figAFM}a shows a MoSe$_{2}$ monolayer on a $\SI{14\pm0.7}{\nano\meter}$ thick hBN multi-layer crystal. Although hardly
visible in the optical microscope image, the MoSe$_{2}$ layer strongly wrinkles in the upper half part of the flake. The bottom half and the region
on the SiO$_{2}$ exhibit bubble formation. This contrast can be attributed to different speeds and contact pressures during the viscoelastic stamping process.

Figure \ref{figAFM}b shows the device structure after an annealing procedure at $\SI{150}{\degreeCelsius}$ for $\SI{20}{\minute}$. We note significant changes in the monolayer's morphology. To further quantify the roughness, we define regions indicated as solid and dashed squares with an area of $\SI{9}{\micro\meter\squared}$ on the hBN region and $\SI{4}{\micro\meter\squared}$ on the SiO$_{2}$, respectively, and calculate the RMS roughness for these areas. The RMS roughness on the SiO$_{2}$ and hBN significantly changes from $\sigma_{\text{SiO}_{2}} = \SI{2.27}{\nano\meter}$ and $\sigma_{\text{hBN}} = \SI{2.35}{\nano\meter}$ to $\sigma'_{\text{SiO}_{2}} = \SI{1.61}{\nano\meter}$ and $\sigma'_{\text{hBN}} = \SI{1.02}{\nano\meter}$ after annealing. The bubble density reduces while smaller bubbles aggregate in bigger bubbles which we attribute to water and polymer residue from the viscoelastic stamping with PDMS under ambient conditions\cite{CastellanosGomez.2014,khestanova2016universal}.

We note, that the left part of the monolayer is shifted by $\SI{2.5}{\micro\meter}$ to the right compared to the as-exfoliated image. This could be explained by the low vapour pressure of the water residue between the different crystals inside the annealing device. The expanding gas leaving the structure lifts parts of the TMDC, while the MoSe$_{2}$ minimises its surface energy on the hBN in the process by moving inwards. Overall, we observe a smoother MoSe$_{2}$ surface on the hBN substrate that is consistent with recent works\cite{khestanova2016universal}.

\section{Peak and linewidth distributions}

In the main text we use combined plots for the peak positions and the linewidths of MoSe$_{2}$ and WSe$_{2}$ in different configurations (Fig. 1d and Fig. 2d).
In order to unfold these plots, here we show the individual distributions for both parameters.

\subsection{MoSe$_{2}$}

\begin{figure}
	\begin{center}
		\includegraphics[width=0.5\textwidth]{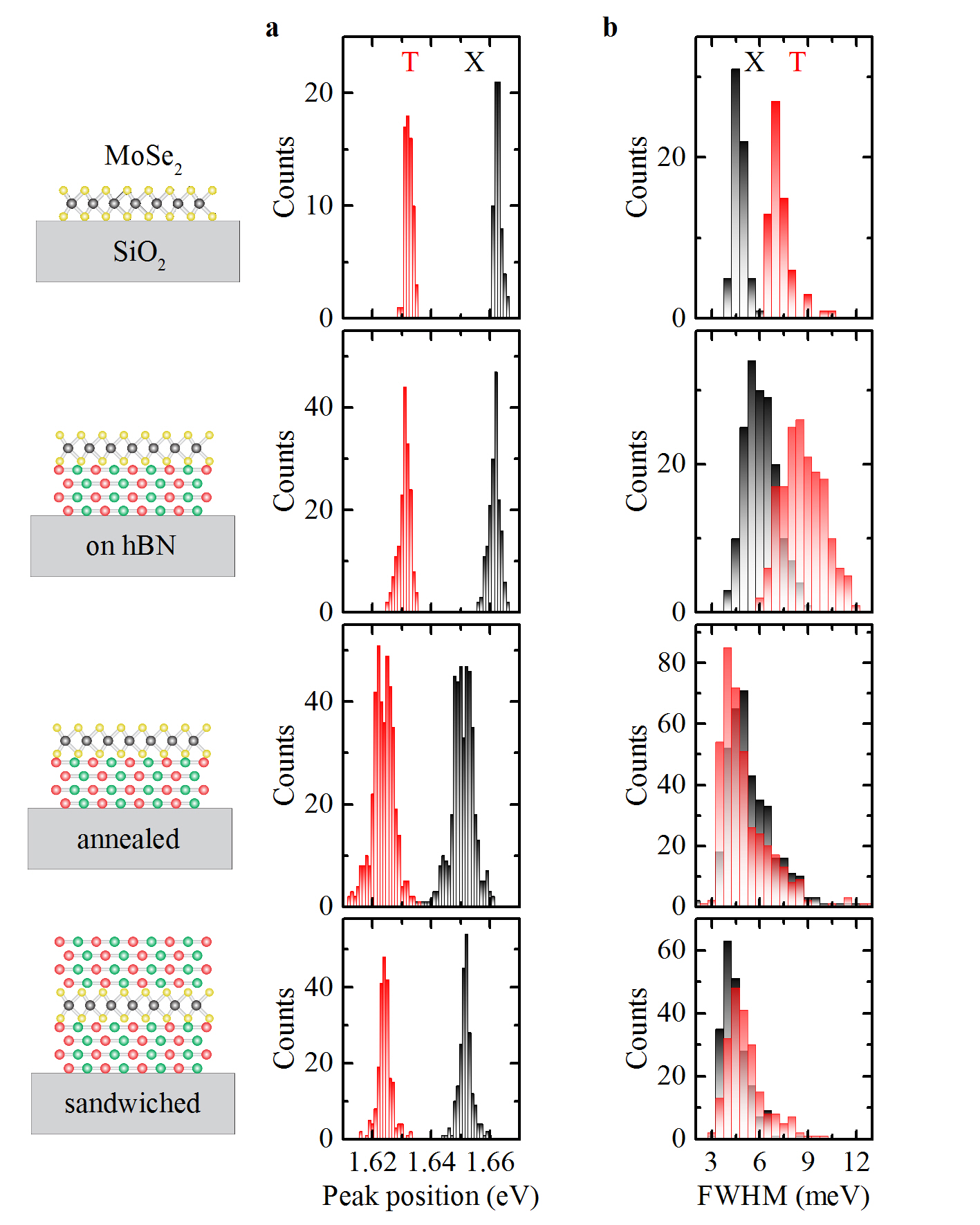}
	\end{center}
	\renewcommand{\figurename}{Fig.}
	\caption{\textbf{MoSe$_{2}$ statistics.}
		MoSe$_{2}$ peak energy and linewidth evolutions for different substrate and capping configurations.
		(a) Trion T (red) and neutral exciton X (black) peak energy distributions. The upper panel shows the fitted peak positions for MoSe$_{2}$ on SiO$_{2}$.
			The central panels show the fitted peak positions of both exciton species for the case of MoSe$_{2}$ directly exfoliated on top of hBN and the positions after an annealing step.
			The lower panel shows the peak energies for MoSe$_{2}$ sandwiched between hBN layers.
		(b) Corresponding linewidth distributions for MoSe$_{2}$ on SiO$_{2}$ (upper panel), MoSe$_{2}$ on hBN (central panels) and the hBN/MoSe$_{2}$/hBN heterostructure (lower panel).
	}
	\label{fig:MoSe2Dist}
\end{figure}

Figure \ref{fig:MoSe2Dist} shows the obtained histograms after fitting of the spectral data for the different MoSe$_{2}$ configurations discussed in the main text. The left panels of Fig.\ref{fig:MoSe2Dist}a show the evolution of the peak positions of the neutral exciton X and the negatively charged exciton T. The corresponding linewidth distributions are presented in Fig.\ref{fig:MoSe2Dist}b.

\newpage

\subsection{WSe$_{2}$}

\begin{figure}
	\begin{center}
		\includegraphics[width=0.6\textwidth]{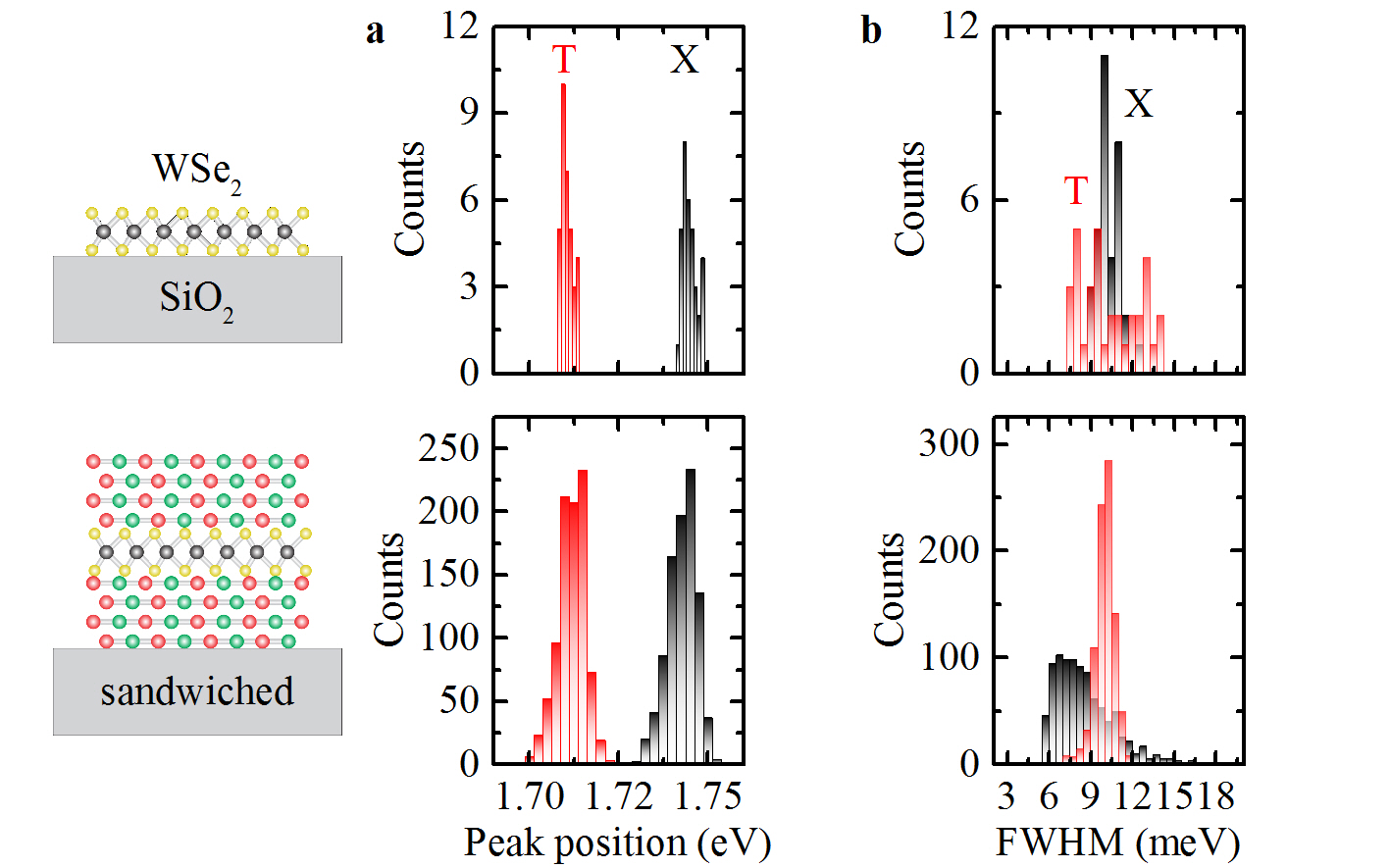}
	\end{center}
	\renewcommand{\figurename}{Fig.}
	\caption{\textbf{WSe$_{2}$ statistics.}
		WSe$_{2}$ peak energy and linewidth evolutions for different substrate and capping configurations.
		(a) Trion T (red) and neutral exciton X (black) peak energy distributions. The upper panel shows the fitted peak positions for WSe$_{2}$ on SiO$_{2}$.	The lower panel shows the peak energies for WSe$_{2}$ sandwiched between hBN layers.
		(b) Corresponding linewidth distributions for WSe$_{2}$ on SiO$_{2}$ (upper panel) and the hBN/WSe$_{2}$/hBN heterostructure (lower panel).
	}
	\label{fig:WSe2Dist}
\end{figure}

Figure \ref{fig:WSe2Dist} shows the peak positions (a) and linewidths (b) distributions for the WSe$_{2}$/SiO$_{2}$ and hBN/WSe$_{2}$/hBN configurations.
\newline

\subsection{Statistics summary}

For clarity, we list all obtained statistical relevant values from the main text in table \ref{tab:EandW}.

\begin{center}
\begin{table}[h!]
	\renewcommand{\tablename}{Table}
\begin{tabular}{|c||c|c|c|c|}
	\hline 
	& \thead{Neutral exciton\\peak energy (meV)} & \thead{Neutral exciton\\linewidth (meV)} & \thead{Trion\\peak energy (meV)} & \thead{Trion\\linewidth (meV)} \\ 
	\hhline{|=|=|=|=|=|} 
	\thead{ MoSe$_{2}$/SiO$_{2}$ }		& $1663.1\pm1.2$ & $ 5.0\pm0.5$ & $1631.8\pm1.3$ & $7.0\pm0.8$ \\ 
	\hline 
	\thead{ MoSe$_{2}$/hBN }			& $1661.8\pm2.0$ & $ 6.3\pm1.0$ & $1630.3\pm2.0$ & $8.4\pm1.3$ \\ 
	\hline 
	\thead{ MoSe$_{2}$/hBN (annealed) }	& $1651.0\pm3.9$ & $ 5.7\pm1.5$ & $1623.0\pm3.7$ & $4.8\pm1.5$ \\ 
	\hline 
	\thead{ hBN/MoSe$_{2}$/hBN }		& $1652.1\pm2.3$ & $ 4.7\pm0.9$ & $1623.6\pm2.4$ & $4.9\pm1.3$ \\ 
	\hline 
	\thead{ WSe$_{2}$/SiO$_{2}$ } 		& $1744.8\pm2.0$ & $10.3\pm0.7$ & $1710.4\pm1.6$ & $10.1\pm2.1$ \\ 
	\hline 
	\thead{ hBN/WSe$_{2}$/hBN }			& $1743.8\pm4.8$ & $ 9.8\pm1.4$ & $1710.5\pm3.8$ & $ 8.4\pm1.9$ \\ 
	\hline 
	\thead{ MoS$_{2}$/SiO$_{2}$ }		& $1947.4\pm0.3$ & $14.7\pm0.7$ & $1910.7\pm0.3$ & $23.4\pm0.8$ \\
	\hline
	\thead{ hBN/MoS2/hBN }				& $1955.8\pm0.5$ & $ 4.8\pm1.0$ & $1926.2\pm0.5$ & $ 6.8\pm0.9$ \\
	\hline
\end{tabular}
\caption{Summarised peak energies and line widths of the different material combinations discussed in the main text.}
\label{tab:EandW}
\end{table}
\end{center}
\newpage 

\bibliographystyle{apsrev4-1}
\bibliography{SuppBibliography}